\documentclass[a4paper,11pt]{article}
\usepackage{jheppub} % for details on the use of the package, please
\usepackage{epsfig,multicol} 
\usepackage{amsmath}
\usepackage{amssymb}
\usepackage{subfigure}
\usepackage{here}
\usepackage{wrapfig}
\usepackage{graphicx}
\usepackage[T1]{fontenc} % if needed

\def\eq#1{{Eq.~(\ref{#1})}}

\def \to  {\rightarrow}
\def \be {\begin{equation}}
\def \ee {\end{equation}}
\def \lf  {\left(}
\def \rt  {\right)}
\def \L  {\left[}
\def \R  {\right]}
\def \pa  {\partial}

\def \Gm  {\Gamma}
\def \f   {\frac}
\def \s   {\sqrt}
\def \i   {\int}
\def \o   {\omega}

\title{\boldmath Schwinger Pair Production in Hot Anti-de Sitter Space}

\author[a]{Prasant Samantray}
\author[b]{and Suprit Singh}

\affiliation[a]{Department of Physics, BITS-Pilani Hyderabad Campus, Jawahar Nagar, Shamirpet Mandal, Secunderabad 500078, India}
\affiliation[b]{Department of Mathematics and Statistics, University of New Brunswick, Fredericton, E3B 5A3 Canada}

\emailAdd{prasant.samantray@hyderabad.bits-pilani.ac.in}
\emailAdd{suprit.singh@unb.ca}

\abstract{We consider particle production in $1+1$ dimensional thermal Anti-de Sitter space under the influence of a constant electric field. The vacuum-persistence amplitude is given by a non-relativistic tunnelling instanton once we interpret the system as being governed by an ``equivalent'' non-relativistic Schr\"odinger equation. Working in the WKB approximation, we calculate the tunnelling rate in anti de Sitter space at finite temperature and observe that the particle production rate is enhanced. Additionally, it is observed that there is a critical temperature beyond which the production rate is affected by the thermal environment. We claim this to be a new result for Anti-de Sitter space in the semi-classical approximation.}

\begin{document} 
\maketitle
\flushbottom

\section{Introduction}
\label{sec:intro}

In the ``Consequences of Dirac's theory of positrons'' \cite{Heisenberg:1935qt}, Heisenberg and Euler suggested that quantum vacuum fluctuations can be uplifted to ``real'' observable pair of particles if given an external electric field of sufficient strength. To the leading order, this pair production rate per unit volume in flat  spacetime at zero temperature is 
\be
\label{rate}
\Gamma \sim \exp\left({-\frac{\pi m^2}{q\,E}}\right);
\ee
a result in quantum electrodynamics first computed by Julian Schwinger \cite{Schwinger:1951a}. These produced pairs constitute a current that backreacts to destabilize the otherwise classically stable background electric field configuration. This effect named after Schwinger has been studied extensively in the literature (see ref.~\cite{Dunne:2004nc} for a review) and active searches are going on to observe the pair production with high intensity lasers in laboratories \cite{Dunne:2010}. Schwinger mechanism also plays a key role in the study of quantum fields in curved spacetime and gains a multitude of features with gravity acting on the quantum fields in tandem with the electric field. Such a setting is natural, for example, in the quasi-de Sitter background during inflation (in the context of inflationary magnetogenesis, nucleation of strings, membranes, domain walls, and monopoles) \cite{inflationary, nucleation}, and in pair creation due to charged black holes \cite{Gibbons}. Schwinger effect has also been studied in Anti-de Sitter (AdS) space \cite{dsads}, and recently, forming an indispensable piece in relating the ``cosmic censorship'' conjecture with the ``weak gravity'' conjecture \cite{Crisford}. 

The particle production rate can be computed using different formalisms and each has its own interpretation of the effect. The one relevant for this work is reducing the original field-theoretic problem to an equivalent tunnelling problem in point-particle quantum mechanics. The production rate is then given by the Euclidean action computed over the tunnelling instanton, or equivalently, a WKB quantum tunnelling probability \cite{Affleck:1982, Affleck:1982a, Stephens:1988, Stephens:1989, Srinivasan:1999} to go across a barrier. At zero temperature, this process is solely dictated by quantum tunnelling and the energy extracted from the background electric field. However, in considering a thermal Minkowski background, the effects of finite temperature and thermal fluctuations are seen to enhance the pair creation process by also extracting energy from the heat bath \cite{AdamB}. Consequently, there is a critical temperature $T_c$ where a first order phase transition takes place such that below $T_c$ the effect is completely quantum dominated, and for $T>T_c$ thermal fluctuations enhance the production rate. Anti-de Sitter space presents a similar case with a critical threshold field, $E_c$ below which there is no particle production at all due to the confining nature of the AdS. 

In this note, we consider particle production in $(1+1)$ dimensional ``thermal'' AdS space under the influence of a constant electric field. We use the instanton techniques as described above by mapping the problem to a non-relativistic quantum tunneling problem. Working in the WKB approximation, we calculate the tunneling rate in AdS space at finite temperature and observe that the particle production rate is enhanced similar to what is seen in thermal Minkowski background with a critical temperature beyond which the production rate is affected by the thermal environment. As a precursor, we first illustrate the technique in flat background in Sec.~\ref{sec:flat} before applying it to AdS (Sec.~\ref{sec:AdS}).

\section{Particle Production in Minkowski Space}
\label{sec:flat}

\subsection{At Zero Temperature}
\label{sec:flat space at zero T}

\noindent In $(1+1)$ dimensional Minkowski space $\{\mathbb{R}^{1,1}: (T,X)\}$, the electromagnetic field tensor has only one independent component and is, therefore, proportional to the alternating tensor:~$F_{\mu\nu} = - E \sqrt{-g} \,\epsilon_{\mu\nu}$, where $E>0$ is a constant to be identified with the electric field acting on the charged quanta. A suitable choice of gauge potential for such a field is given by $A_0 = E X$ with foresight that the resulting dynamical equation at the end will be time independent. We consider a massive charged scalar field propagating on the flat background coupled to the gauge field given by,
\be
\f{1}{\sqrt{-g}} \lf  \pa_{\mu} - i q A_{\mu} \rt \L \s{-g} g^{\mu\nu} \lf  \pa_{\nu} - i q A_{\nu} \rt  \R \Phi = M^2 \Phi, \label{MasterEq}
\ee
\noindent where $q>0$. Since the metric and the choice of gauge potential is static in nature, the above equation admits a solution of the form $\Phi(X,T)= \psi(X) \exp({-i\o T})$ with $\o \geq 0$ so that we have for $\psi(X)$:
\be
-\f{1}{2}  \f{d^2 \psi}{dX^2}  + \f{1}{2} \L M^2 - \lf \o + qEX \rt^2 \R \psi = 0 \label{SchroEq-Mink}.
\ee
The above equation is exactly a time-independent Schr\"odinger equation determining the wave function of a particle of unit mass moving in the potential,
\be 
\label{potpflat}
V_{+}(X) = \f{1}{2} \L M^2 - \lf \o + qEX \rt^2 \R,
\ee
\begin{figure}[h]
\quad\includegraphics[scale=0.49]{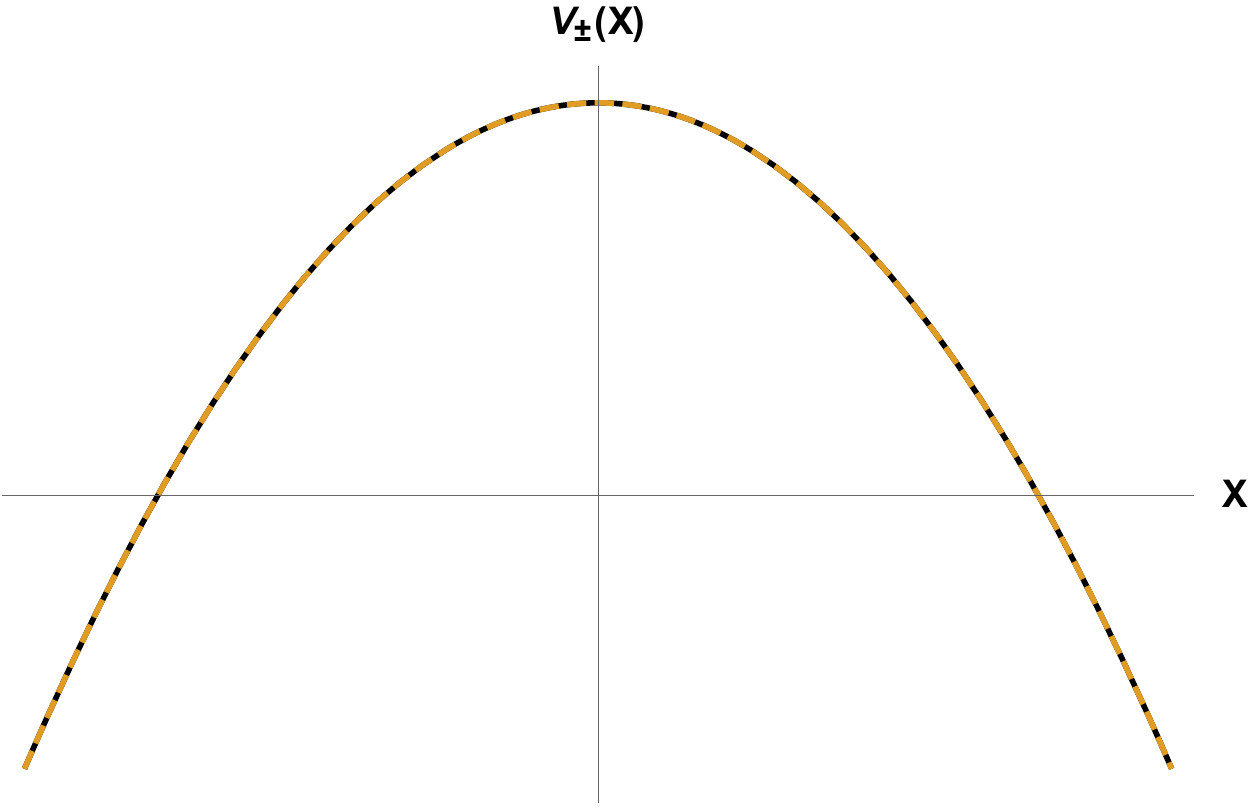}
\includegraphics[scale=0.50]{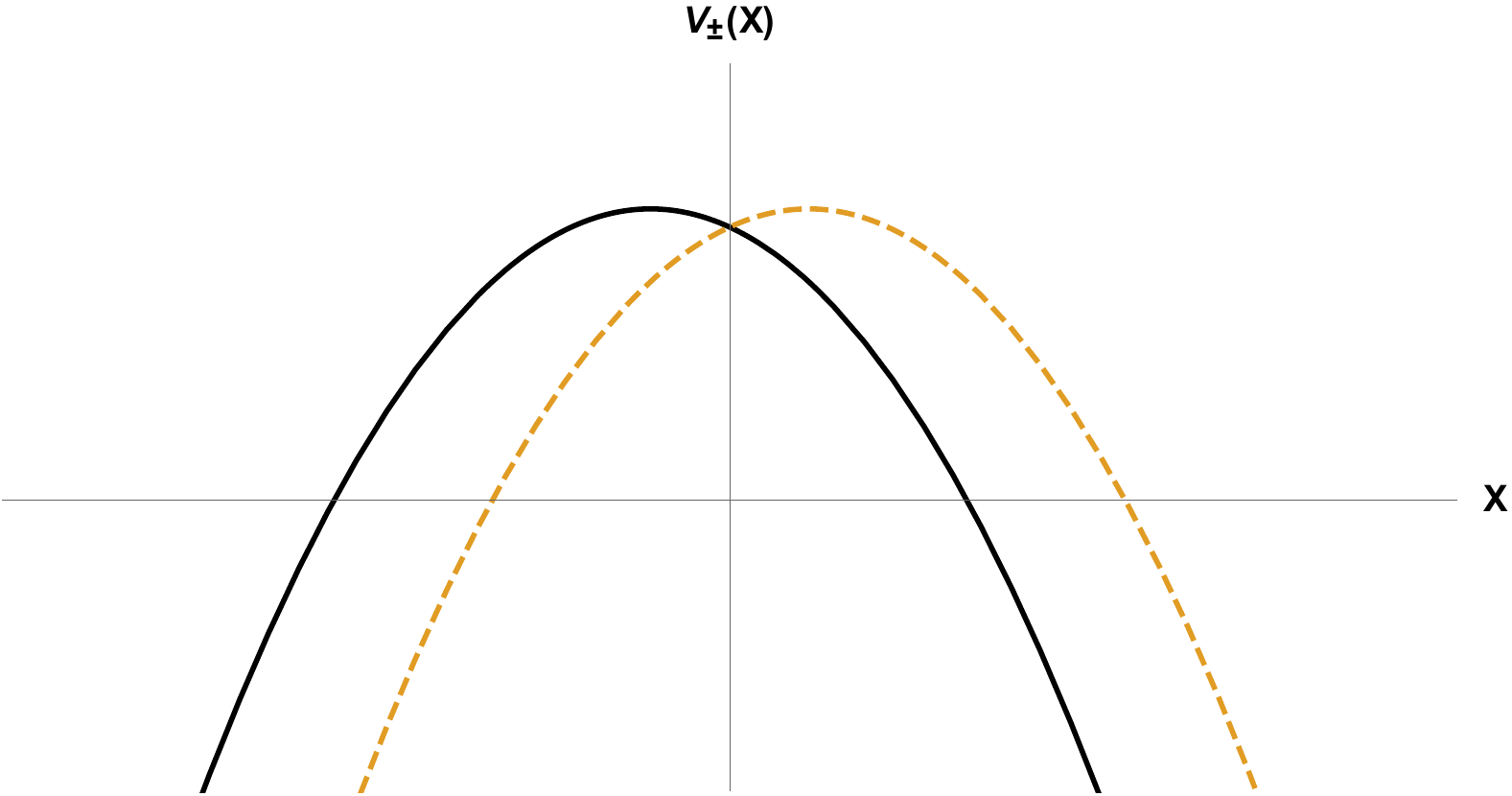}
\caption{The tunneling potentials for the (scalar) electron and positron in flat spacetime at zero temperature (left) and finite temperature (right). We see that in the case of finite temperature, with $\omega\neq0$, an asymmetry exists in the potential which shifts it away such that the point of nucleation ceases to be the point of symmetry.}
\end{figure}

with vanishing energy eigenvalue. This suggests an interpretation of the process as tunneling of a positively charged mode (identified with the positron). The negatively charged mode (or the electron) also tunnels through a potential barrier obtained by replacing $q \to -q$ to get
\be
\label{potmflat}
V_{-}(X) = V_+(X)\big|_{q\to -q} = \f{1}{2} \L M^2 - \lf \o - qEX \rt^2 \R. 
\ee
In this instanton picture, the virtual particle-antiparticle pair is spontaneously created at a location where $V_{+}(X) = V_{-}(X)$ which is at $X=0$ and tunnel through the respective potentials to become a real pair. The ``distance'' a particle has to traverse through the barrier is given by the turning point of the barrier. For instance, the positron needs to tunnel a distance $X_0 = (M - \o)/{qE}$. Considering completely static solutions to \eq{MasterEq} by taking $\o = 0$, the WKB tunneling amplitude for the positron is given by
\begin{align}
{\cal A}_{+} =& \exp \L -\i^{X_0}_{0} dX \,\s{2\,V_{+}(X)}\R \nonumber \\
=& \exp \L -\f{\pi M^2}{4\,qE} \R \label{Zero-T-Mink}
\end{align}
The probability for positron to tunnel through is simply the amplitude squared and hence  $\Gm_{+} = |{\cal A}_{+}|^2 = \exp \L -\pi M^2/2qE \R$. For $\o=0$, the electron tunnels through a barrier which is just a mirror image of the positron potential. We, thus, have $\Gm_{-} = |{\cal A}_{-}|^2 = \exp \L -\pi M^2/2qE \R$. Therefore, the total probability of pair creation is just,
\be
\Gm = \Gm_{+} \Gm_{-} = \exp \L -\f{\pi M^2}{qE} \R,
\ee
which is the correct leading order rate for zero temperature Schwinger mechanism apart from the pre-factors. 

\subsection{At Finite Temperature}
\label{sec:flat space at non-zero T}

\noindent In the presence of a thermal bath, the virtual particle can borrow a certain amount of energy from the bath and move up the potential barrier. As a result, it needs to tunnel through a ``smaller'' distance as compared to the zero temperature case. For this we take, $0< \o < M$, where this assumption ensures that thermal fluctuations never fully dominate quantum tunneling as the charged quanta can never acquire energy greater than its rest mass. With the same potentials as in the previous case, that is, \eq{potpflat} but with $\o\neq 0$, the WKB tunneling amplitude for the positron with $X_0 = (M - \o)/qE$ is given by 
\begin{align}
{\cal A}_{+} =& \exp \L -\i^{X_0}_{0} dX \,\s{2\,V_{+}(X)}\R \nonumber \\\nonumber \\
=& \exp \L -\f{M^2}{2qE} \arccos \lf \f{\o}{M}\rt + \f{\o \s{M^2 - \o^2}}{2qE} \R,
\end{align}
and corresponding quantum tunneling probability for the positron is given by  $|{\cal A}_{+}|^2$. However, there is also another probability associated with the process that the particle excitation extracts out the energy ``$\o$'' from the bath which is given by the Boltzmann factor. Therefore, the total probability which is (quantum + thermal) for the positron to tunnel through is given by 
\be
\Gm_{+} = \exp \L -\beta \o - \f{M^2}{qE} \arccos \lf \f{\o}{M}\rt + \f{\o \s{M^2 - \o^2}}{qE} \R, 
\ee 
where $\beta=1/T$ is the inverse temperature. The counterpart of this for an electron is the same, that is, $\Gm_{-} = \Gm_{+}$ by symmetry arguments. Therefore, the total probability for the pair creation is given by
\be
\label{ratefinitetemp}
\Gm = \Gm_{+} \Gm_{-} = \exp \L -2\beta \o - \f{2M^2}{qE} \arccos \lf \f{\o}{M}\rt + \f{2\o \s{M^2 - \o^2}}{qE} \R 
\ee
We need to fix one last thing, that is, how much energy does the pair extract from the bath? This is given by the condition that the probability be maximized with respect to the energy extracted from the bath, that is,
\be
\label{ofinitetemp}
\f{\pa\Gm}{\pa \o} = 0~~\implies~\o = M \s{1 - \f{T^2_c}{T^2}}\,;~~T_c = \f{qE}{2M}.
\ee
This implies for temperatures $T<T_c$ the process is solely dictated by quantum fluctuations and the heat bath has no effect on the probability. However, for $T>T_c$, there is a phase transition in the decay probability and the rate is given by \eq{ratefinitetemp} along with \eq{ofinitetemp} such that
\be
\Gm_{T>T_c} = \exp \L -\f{2 M}{T_c} \arcsin \lf \f{T_c^2}{T^2}\rt - \f{M}{T} \sqrt{1-\f{T_c^2}{T^2}} \,\R 
\ee
enhancing the pair production. With this setup in place, we are now in position to replace flat spacetime with the AdS background.

\section{Particle Production in Anti-de Sitter Space}
\label{sec:AdS}

We consider anti-de Sitter space in the global, static coordinate system given by the metric: 
\be
ds^2 = - \lf 1 + x^2/l^2 \rt dt^2 + \f{dx^2}{\lf 1 + x^2/l^2  \rt}
\ee
which covers the entire manifold with no pathological issues. We choose to work in the static gauge again with the potential $A_0 = E x$ and solve \eq{MasterEq} taking an ansatz of the form
\be
\Phi(x,t) = \f{\psi(x)}{\s{1 + x^2/l^2}}\,e^{-i\o t}
\ee 
so that the corresponding equation for the positron is,
\be
-\f{1}{2}  \f{d^2 \psi}{dx^2}  + \f{M^2 \lf 1 + x^2/l^2 \rt + 1/l^2 - \lf \o + qEx \rt^2}{2\lf 1 + x^2/l^2 \rt^2} \psi= 0. \label{SchroEq-AdS}
\ee
This is again equivalent to the time-independent Schr\"odinger equation for a point-particle of unit mass with vanishing energy eigenvalue, tunneling through a potential barrier given by 
\be
V_{+} (x) =  \f{M^2 \lf 1 + x^2/l^2 \rt + 1/l^2 - \lf \o + qEx \rt^2}{2\lf 1 + x^2/l^2 \rt^2}
\label{AdSpotential}
\ee
provided $0\leq \o< M$. The potential barrier for the electron, $V_{-}(x)$, is obtained by the usual replacement of $q \to -q$. As noted previously, the virtual pair is instantaneously created at the location where $V_{-}(x) = V_{+}(x)$, that is, at $x=0$ following which the particles tunnel through their respective barriers. 

\subsection{At Zero Temperature}
\label{sec:anti de Sitter space at zero T}
 
Keeping the analysis in line with that of previous section that had flat spacetime, we look at the solutions with $\o = 0$ for the AdS background with zero temperature. Considering first the tunneling of a positron at zero temperature, we see that the turning point for $V_{+}(x)$ is at 
\be
x_0 = \f{\s{1 + M^2 l^2}}{\s{q^2 E^2 l^2 - M^2}}.
\ee
With this, the WKB amplitude is given by   
\begin{eqnarray}
{\cal A}_{+} &=& \exp \L -\i^{x_0}_{0} dx\,\s{2 V_{+}(x)}\R  \nonumber \\
&=& \exp \L-\f{\pi l}{2} \left( \s{q^2E^2l^2 + 1/l^2} - \s{q^2E^2l^2 - M^2}\right) \R. \label{Zero-T-AdS}
\end{eqnarray}
Again, by virtue of the symmetry in the potentials, that is $V_{+}(x)=V_{-}(x)$ for $\o=0$, the total probability for the pair creation is given by, 
\begin{eqnarray}
\Gm &=& |{\cal A}_{+}|^2 |{\cal A}_{-}|^2 \nonumber \\
&=& \exp \L-2\pi l \left( \s{q^2E^2l^2 + 1/l^2} - \s{q^2E^2l^2 - M^2}\, \right)\R \nonumber \\
&\simeq & \exp \L-2\pi l^2 \left( qE - \s{q^2E^2 - M^2/l^2}\, \right)\R\nonumber\\
&=& \exp \L-2\pi l^2 \left( qE - \s{q^2E^2 - q^2E_c^2 }\, \right)\R \label{Zero-T-AdS-Probability}
\end{eqnarray}
where in the penultimate step we assumed $q^2E^2l^4\gg1$. The result matches the tunneling probability reported previously in the literature. The effect of replacing flat spacetime with AdS brings in a critical electric field $qE_c = M/l$ below which there is no particle production at all.

\subsection{At Finite Temperature}
\label{sec:anti de Sitter space at non-zero T}

We now consider a thermal anti de Sitter background which according to the previous analysis implies that we need to take $0<\o <M$ in the respective tunneling potentials. With the potential \eq{AdSpotential}, the quantum WKB tunneling amplitude for the positron is given by
\begin{eqnarray}
\label{probintadsT}
{\cal A}_{+} &=& \exp \L -S_+\R = \exp \L -\i^{x_0}_{0} dx\,\s{2 V_{+}(x)}\R  \nonumber \\
&=& \exp \L -\i^{x_0}_0 \f{dx}{\lf 1 + x^2/l^2 \rt}\, \s{M^2 \lf 1 + x^2/l^2 \rt + 1/l^2 - \lf \o + qEx \rt^2}\,\R.
\end{eqnarray}
where the turning point is 
\be
x_0 = \f{-\o qE l^2 + M l \s{q^2 E^2 l^2 - M^2 + \o^2}}{q^2 E^2 l^2 - M^2}.
\ee
It is helpful to define the variables,
\begin{align}
a &= \frac{M \sqrt{-\frac{M^2}{l^2}+\frac{\o^2}{l^2}+q^2E^2} - qE \o}{l \left(q^2E^2-\frac{M^2}{l^2}\right)};\qquad  b = \frac{M \sqrt{-\frac{M^2}{l^2}+\frac{\o^2}{l^2}+q^2E^2} + qE \o}{l \left(q^2E^2-\frac{M^2}{l^2}\right)}; \nonumber \\
\gamma &= \sqrt{\f{a}{b}};~~\alpha = \sqrt{\f{1 + a^2}{1 + b^2}} ;~~\beta = \sqrt{\frac{a b +\sqrt{\left(a^2+1\right) \left(b^2+1\right)}-1}{2 \left(b^2+1\right)}}; ~~\sigma = \sqrt{\alpha - \beta^2}\nonumber \\
I_1 &= -2 \arctan(\gamma); \nonumber \\
I_2 &= \f{1 + \alpha}{2\sigma} \left[\arctan\left(\f{\gamma - \beta}{\sigma}\right) +  \arctan\left(\f{\gamma + \beta}{\sigma}\right)\right]; \nonumber \\
I_3 &= \f{1 - \alpha}{4 \beta} \left[\log \left[\f{\gamma^2 - 2\beta \gamma + \alpha}{\gamma^2 + 2\beta \gamma + \alpha} \right] \right], \nonumber
\end{align}
to express the result of the integral in \eq{probintadsT}:
\begin{equation}
S_{+} = l^2 \sqrt{\left(q^2E^2-\frac{M^2}{l^2}\right)} \left[I_1 + I_2 + I_3 \right]. \label{AdS-NonzeroT-Exact-Answer}
\end{equation}
The above result is completely consistent in the sense that it reproduces (i) Eq (\ref{Zero-T-Mink}) in the limit $\o \rightarrow 0;~l \rightarrow \infty$, (ii) Eq (\ref{ratefinitetemp}) in the limit $\o \neq 0;~l \rightarrow \infty$, and (iii) Eq (\ref{Zero-T-AdS}) in the limit $\o \rightarrow 0;~0 <l <\infty$. Following the analysis in the previous section, we now introduce the Boltzmann factor to specify the extracted energy from the heat bath so that the total tunnelling probability for the positron is given by
\be
\Gm_{+} = e^{-\beta \o - 2 {\cal S}_{+}} \nonumber
\ee
By symmetry of the problem, the electron tunnels with the same probability so that $S_{-} = S_{+}$. Therefore the total pair creation probability is given by 
\be
\Gm = \Gm_{+}\Gm_{-} =  e^{-2 \beta \o - 4 {\cal S}_{+}} \label{AdSProbability}
\ee
The energy extracted from the heat bath to thermally assist the process of tunnelling is given by the condition that the above probability be extremal, that is, $\pa_\o\Gm = 0$. However, it turns out that this condition leads to a complicated transcendental equation for $\o$ which requires numerical analysis. Therefore, to proceed analytically and in order to gain a qualitative insight, we expand the result of \eq{AdS-NonzeroT-Exact-Answer} in the large AdS scale approximation i.e., we assume that the AdS scale ``$l$'' is larger than any other existing length scale in the theory. Under such an approximation, we get

\begin{eqnarray} 
S_{+} &=& \frac{6 M^4 \arctan \left(\sqrt{\frac{M - \o}{M + \o}}\right) + \o \left(2 \o^2-5 M^2\right) \sqrt{(M^2 - \o^2}}{24 l^2 q^3E^3} \nonumber \\
&~& - ~~ \frac{\left[\o \sqrt{(M^2 - \o^2}-2 M^2 \arctan \left(\sqrt{\frac{M - \o}{M + \o}}\right)\right]}{2 qE} + {\cal O}\left(\frac{1}{l^3}\right) + ...
\end{eqnarray} 

The above expression in conjunction with the condition (\ref{AdSProbability}) yields

\be
\f{2 \xi}{qE} + \f{2 \xi^3}{3q^3E^3l^2} = \beta;~~~\xi = \s{M^2 - \o^2}.
\ee
The above cubic equation has the solution: 
\be
\o^2 = M^2-\frac{\left(\sqrt[3]{2} \left(q^3 E^3 l^2 \left(3 \beta +\sqrt{9 \beta ^2+16 l^2}\right)\right)^{2/3}-2\ 2^{2/3} q^2 E^2 l^2\right)^2}{4 \left(q^3 E^3 l^2 \left(3 \beta +\sqrt{9 \beta ^2+16 l^2}\right)\right)^{2/3}}. \label{EnergyExtractedAdS}
\ee
However, we need to satisfy the condition $0<\o < M$ at all times which gives an additional constraint 
\be
T> T_c~~;~~~T_c = \f{3 q^3 E^3 l^2}{2M(M^2 + 3 q^2 E^2 l^2)} \label{PhaseTransitionAdS}
\ee
This is a new result in anti-de Sitter space in the semi-classical approximation, and the temperature has an effect on the particle production only beyond a certain critical value similar to what we had for flat spacetime. Comparing this with Eq (\ref{ofinitetemp}), we observe that in anti-de Sitter space, temperature starts affecting the pair production rate at a lower critical temperature as compared to flat spacetime. Therefore the effect of curvature is to bring down the temperature threshold for enhancement of pair production. The total probability of pair production in hot anti-de Sitter space is given by the set of Eqs (\ref{AdS-NonzeroT-Exact-Answer}), (\ref{AdSProbability}), (\ref{EnergyExtractedAdS}), with the condition in (\ref{PhaseTransitionAdS}). In the event that $0<T<T_c$, the probability is given by Eq (\ref{Zero-T-AdS-Probability}).

\section{Discussion and Summary}
\label{sec:discussion}

In this paper we considered the issue of charged pair production in thermal anti-de Sitter space. Though the mathematical results derived herein are self-explanatory, few important comments are in order. To begin with, the literature is unclear as to how non-relativistic quantum mechanics concretely arises from a relativistic theory (see \cite{Padmanabhan} and references therein). Our approach in this paper is based on the crucial interpretation that an equivalent Schr\"odinger equation can be used as a proxy for the relativistic Klein-Gordon equation. Though conceptually speculative, the interesting fact is that such an approach precisely recovers the results for the Schwinger mechanism in both Minkowski (in the absence/presence of a thermal bath) as well as in anti de Sitter space (for zero temperature). Taking this forward we derived a novel result in hot anti-de Sitter space. However, we would like to state that a full blown quantum field theory calculation using thermal field theory in anti de Sitter space would be a complete exercise. We reserve this issue for our future work.

Secondly, in this paper we limited ourselves to a charged scalar field with a mass such that $M^2 > 0$. In anti-de Sitter space it well known that a scalar field with $M^2 < 0$ albeit satisfying the Breitenlohner-Freedman bound \cite{Breitenlohner:1982}, is a stable solution. At this point it is unclear as to how would our approach address this. Even in the absence of any temperature, a charged scalar with $M^2 < 0$ would probably lead to complications as can be seen from Eq (\ref{AdSProbability}) i.e., a naive replacement of $M^2 \rightarrow -M^2$ leads to nonsensical results with probability violation. Therefore, care needs to be taken for such special cases.

Additionally, we would like to clarify that mapping a relativistic field theoretic problem to a tunnelling calculation in a non-relativistic quantum mechanics crucially depends on the coordinate system being employed. To that end, it is clear that if we had employed Rindler coordinates instead of global Minkowski ones, the potential barrier would have been different. However, the issue with choosing such a coordinate system would have been the pathologies associated with spacetime points where the coordinate system breaks down. In such circumstances the whole tunnelling picture becomes obscure, and is unclear as to how to set up the problem using our approach. That said, the dependence of Schwinger process on coordinate systems is an unresolved issue.

Lastly, by virtue of the AdS/CFT correspondence it would be interesting to derive our results for anti-de Sitter space using the boundary CFT. A qualitative picture can be constructed as follows. Since we restricted ourselves to $1+1$ dimensions, the boundary CFT would have operators that depend only on the time-coordinate. It is well known that the dual operator corresponding to a gauge field $A_{\mu}$ in the bulk anti-de Sitter space is the current ${\cal J^{\mu}}$ in the CFT. In our case since the CFT operators are purely time-dependent, the current has the form ${\cal J^{\mu}}(t)$. In our present work we took a gauge potential such that $A_1=0;~A_0 = Ex$. Therefore, the boundary value of $A_0$ depends on the large cut-off for the global AdS radial coordinate $``x"$. The corresponding CFT operator is then ${\cal J}^0(t)$. It is well known that in any quantum theory which has a time-dependent non-static current/source, particles will be produced with the probability $\Gamma \sim e^{- \int {|{\cal J}^0(\o)|^2} d\o} $, where ${\cal J}^0(\o)$ is the Fourier transform of ${\cal J}^0(t)$. However, to holographically compute the finer details of particle production in the absence/presence of a temperature, one of course needs the full CFT. We hope to make some progress regarding this in our future work.

\section{Acknowledgments}

Research of P.S. is partially supported by CSIR grant 03(1350)/16/EMR-II Govt. of India, and also by the OPERA fellowship from BITS-Pilani, Hyderabad Campus. Research of S.S. is partially supported by the Natural Sciences and Engineering Research Council of Canada.

\end{document}